\newcommand{\beq}{\begin{equation}}
\newcommand{\eeq}{\end{equation}}
\newcommand{\beqn}{\begin{eqnarray}}
\newcommand{\eeqn}{\end{eqnarray}}
\newcommand{\dd}{\mbox{d}}
\newcommand{\dds}{\mbox{\scriptsize d}}
\newcommand{\dps}{\mbox{dPS}}
\newcommand{\smc}{\scriptsize}
\newcommand{\porder}[1]{\mbox{${\cal O}(#1)$}}

\newcommand{\ifdraft}{nodraft}

\newcommand{\ifprep}{false}
\renewcommand{\ifprep}{true}

\newcommand{\psinr}{97--20}

\newcommand{\shiftcaption}{\vspace{-0.3cm}}

\newcommand{\dgpicture}[2]{
\begin{picture}(#1,#2)   
\thicklines
{\ifthenelse{\equal{\ifdraft}{draft}}%
{
}%
{}%
}
}

\newcommand{\fullline}{
\unitlength0.4mm
\begin{picture}(13,4)
\linethickness{0.3mm}
\put(-1,2.0){\line(1,0){15}}
\thinlines
\end{picture}
}

\newcommand{\dashline}{
\unitlength0.4mm
\begin{picture}(20,4)
\linethickness{0.3mm}
\put(-1,2.0){\line(1,0){4}}
\put(8,2.0){\line(1,0){4}}
\put(17,2.0){\line(1,0){4}}
\thinlines
\end{picture}
}

\newcommand{\dotline}{
\unitlength0.4mm
\begin{picture}(9,4)
\linethickness{0.3mm}
\put(-1,2.0){\line(1,0){1}}
\put(4,2.0){\line(1,0){1}}
\put(9,2.0){\line(1,0){1}}
\thinlines
\end{picture}
}

\newcommand{\longdashline}{
\unitlength0.4mm
\begin{picture}(22,4)
\linethickness{0.3mm}
\put(-1,2.0){\line(1,0){10}}
\put(13,2.0){\line(1,0){10}}
\thinlines
\end{picture}
}

\newcommand{\dashdotline}{
\unitlength0.4mm
\begin{picture}(17,4)
\linethickness{0.3mm}
\put(-1,2.0){\line(1,0){5}}
\put(8,2.0){\line(1,0){1}}
\put(13,2.0){\line(1,0){5}}
\thinlines
\end{picture}
}

\documentstyle[12pt,citesort,epsfig,ifthen]{article}

\catcode`\@=11
\long\def\@makefntext#1{
\protect\noindent \hbox to 3.2pt {\hskip-.9pt  
$^{{\ninerm\@thefnmark}}$\hfil}#1\hfill}		

\def\@makefnmark{\hbox to 0pt{$^{\@thefnmark}$\hss}}  
	
\def\ps@myheadings{\let\@mkboth\@gobbletwo
\def\@oddhead{\hbox{}
\rightmark\hfil\ninerm\thepage}   
\def\@oddfoot{}\def\@evenhead{\ninerm\thepage\hfil
\leftmark\hbox{}}\def\@evenfoot{}
\def\sectionmark##1{}\def\subsectionmark##1{}}

\renewcommand{\thefootnote}{\fnsymbol{footnote}}

\newcounter{appendixc}
\newcounter{subappendixc}[appendixc]
\newcounter{subsubappendixc}[subappendixc]

\renewcommand{\appendix}[1] {\vspace*{0.6cm}
        \refstepcounter{appendixc}
        \setcounter{figure}{0}
        \setcounter{table}{0}
        \setcounter{equation}{0}
        \renewcommand{\thefigure}{\Alph{appendixc}.\arabic{figure}}
        \renewcommand{\thetable}{\Alph{appendixc}.\arabic{table}}
        \renewcommand{\theappendixc}{\Alph{appendixc}}
        \renewcommand{\theequation}{\Alph{appendixc}.\arabic{equation}}
        \noindent{\bf Appendix \theappendixc #1}\par\vspace*{0.4cm}}

\def\abstracts#1{{
	\centering{\begin{minipage}{12.2truecm}\footnotesize\baselineskip=12pt\noindent
	\centerline{\footnotesize ABSTRACT}\vspace*{0.3cm}
	\parindent=0pt #1
	\end{minipage}}\par}} 

\renewenvironment{thebibliography}[1]
	{\begin{list}{\arabic{enumi}.}
	{\usecounter{enumi}\setlength{\parsep}{0pt}
\setlength{\leftmargin 1.25cm}{\rightmargin 0pt}
	 \setlength{\itemsep}{0pt} \settowidth
	{\labelwidth}{#1.}\sloppy}}{\end{list}}

\newcounter{itemlistc}
\newcounter{romanlistc}
\newcounter{alphlistc}
\newcounter{arabiclistc}

\newcounter{sectionc}\newcounter{subsectionc}\newcounter{subsubsectionc}
\renewcommand{\section}[1] {\vspace*{0.6cm}\addtocounter{sectionc}{1} 
\def\@currentlabel{\arabic{sectionc}}
\setcounter{subsectionc}{0}\setcounter{subsubsectionc}{0}\noindent 
	{\normalsize\bf\thesectionc. #1}\par\vspace*{0.4cm}}
\renewcommand{\subsection}[1] {\vspace*{0.6cm}\addtocounter{subsectionc}{1} 
	\setcounter{subsubsectionc}{0}\noindent 
	{\normalsize\it\thesectionc.\thesubsectionc. #1}\par\vspace*{0.4cm}}
\renewcommand{\subsubsection}[1]
{\vspace*{0.6cm}\addtocounter{subsubsectionc}{1}
	\noindent {\normalsize\rm\thesectionc.\thesubsectionc.\thesubsubsectionc. 
	#1}\par\vspace*{0.4cm}}

\renewcommand{\section}[1] {\vspace*{0.6cm}\addtocounter{sectionc}{1} 
\def\@currentlabel{\arabic{sectionc}}
\setcounter{subsectionc}{0}\setcounter{subsubsectionc}{0}\noindent 
	{\normalsize\bf\thesectionc. #1}\par\vspace*{0.4cm}}

\newcommand{\xsection}[1] {
\vspace*{0.6cm}\noindent{\normalsize\bf #1}\par\vspace*{0.4cm}
}     
\newcommand{\fcaption}[1]{
        \refstepcounter{figure}
        \setbox\@tempboxa = \hbox{\footnotesize Fig.~\thefigure. #1}
        \ifdim \wd\@tempboxa > 6in
           {\begin{center}
        \parbox{6in}{\footnotesize\baselineskip=12pt Fig.~\thefigure. #1}
            \end{center}}
        \else
             {\begin{center}
             {\footnotesize Fig.~\thefigure. #1}
              \end{center}}
        \fi}

\newcommand{\tcaption}[1]{
        \refstepcounter{table}
        \setbox\@tempboxa = \hbox{\footnotesize Table~\thetable. #1}
        \ifdim \wd\@tempboxa > 6in
           {\begin{center}
        \parbox{6in}{\footnotesize\baselineskip=12pt Table~\thetable. #1}
            \end{center}}
        \else
             {\begin{center}
             {\footnotesize Table~\thetable. #1}
              \end{center}}
        \fi}

 1
 1
 1

\font\ninerm=cmr9

\begin{document}

{\ifthenelse{\equal{\ifprep}{true}}%
{
\begin{minipage}[t]{10cm}
\footnotesize\it
Talk given at the workshop 
on new trends in HERA physics,\\ 
Ringberg Castle, Tegernsee, 25 - 30 May, 1997
\end{minipage}
\hfill
\hspace*{\fill}
\begin{minipage}[t]{3cm}
\footnotesize
PSI--PR/\psinr\\
August 1997
\end{minipage}
\vspace{1.0cm}
}%
{}%
}

\centerline{\normalsize\bf Deeply Inelastic Hadronic Final States:}
\baselineskip=16pt
\centerline{\normalsize\bf QCD Corrections}
\baselineskip=22pt

\centerline{\footnotesize Dirk Graudenz\footnote{
WWW URL: http://www.hep.psi.ch/graudenz/index.html
}}
\baselineskip=13pt
\centerline{\footnotesize\it Paul Scherrer Institut
}
\baselineskip=12pt
\centerline{\footnotesize\it 5232 Villigen PSI, Switzerland}
\centerline{\footnotesize E-mail: Dirk.Graudenz@psi.ch}

\vspace*{0.9cm}
\abstracts{
The status of the calculation of next-to-leading-order QCD corrections
to hadronic final states in deeply inelastic scattering is reviewed, 
and an overview of the phenomenology (including the measurement of the 
strong coupling constant and the gluon density via jet rates) is given.
We also describe a new universal program ({\tt DISASTER++}) for the calculation 
of (2+1)-jet observables.
}
 
\normalsize\baselineskip=15pt
\setcounter{footnote}{0}
\renewcommand{\thefootnote}{\alph{footnote}}

\section{Introduction}
The electron--proton collider HERA has started to accumulate large luminosity.
This opens up the possibility to study the hadronic final state in deeply 
inelastic scattering to high precision. The motive for this is twofold:
QCD can be tested at large scales in the spacelike regime, and an extraction 
of physical quantities, in particular the running strong coupling constant
$\alpha_s\left(Q^2\right)$ and the gluon density $f_g\left(\xi,Q^2\right)$, 
is possible. The aim of this paper is to give an overview of the status of
next-to-leading-order calculations both for jet-like quantities and for
one-particle-inclusive processes. We will also discuss the phenomenology 
of processes with (2+1) jets\footnote{
Here ``$(n+1)$'' stands for $n$ hard jets and the proton remnant jet.
} 
{} in the final state, and of inclusive particle
spectra.

In the last few years, the technology for the calculation
of QCD corrections in next-to-leading order of perturbation theory
has developed considerably. There are explicit algorithms available
which permit the calculation to be done in a ``universal'' way: the 
infrared singularities are subtracted in such a way that arbitrary 
infrared-safe observables can be calculated numerically. In principle, 
all existing algorithms are variations on a common theme, namely the
interplay of the factorization theorems of perturbative QCD and the 
infrared-safety of the observables under consideration. This will be
reviewed in Section~\ref{IRO}.

The practical 
implementations of this principle can be quite different. In 
Section~\ref{CNLO}, we will discuss the two possible ways to do this:
the phase-space-slicing method and the subtraction method. The extraction
of the singular terms can be done in different ways, and we will describe 
in some detail
a new calculation based on the subtraction formalism and on a general 
partial fractions formula.

An overview of applications for (2+1)-jet-like observables will be given in 
Section~\ref{ANLO}, where we state some results for the cut dependence
of jet cross sections and for transverse momentum spectra. 
One-particle-inclusive processes will be discussed
in Section~\ref{OPIP}. The theoretical predictions 
in the framework of the fragmentation
function picture fail to describe the experimental data
for small values of the photon virtuality $Q^2$
and for small values of the energy~$E_h$ of the 
observed hadron, although there
is excellent agreement at large~$Q^2$ and large~$E_h$. We will give a
semi-quantitative estimate of the region where the fragmentation function 
picture should be applicable. 
The paper closes with a summary and an outlook.

\section{Infrared-safe Observables and Factorization in QCD}
\label{IRO}
Perturbative QCD permits the calculation of processes with partons in the
final state. The gap to experimental data with final-state hadrons
is bridged by means of infrared-safe observables $\cal O$.
Examples are jet cross sections for various jet definitions, and 
event shape variables such as Thrust.
The expectation value of 
an observable $\cal O$ for experimental data is given by
\begin{equation}
\langle {\cal O} \rangle_{\mbox{\smc exp}} = \frac{1}{N} \sum_{I=1}^N 
{\cal O}^{(n_I)}\left(h^{(I)}_1, \ldots, h^{(I)}_{n_I}\right),   
\end{equation}
where $N$ is the number of events, $n_I$ is the number of final-state hadrons
of event $I$, and the $h^{(I)}_i$ are the momentum vectors of the hadrons.
This is to be compared with the expectation value for parton final 
states:
\begin{equation}
\label{thf}
\langle {\cal O} \rangle_{\mbox{\smc th}} =
\sum_n \int \dps^{(n)} \, \sigma^{(n)}\left(p_1, \ldots, p_n\right) \, 
{\cal O}^{(n)}\left(p_1, \ldots, p_n\right).
\end{equation}
Here $n$ is the number of final-state partons with momenta $p_1$, \ldots, 
$p_n$, $\dps^{(n)}$ is the phase space measure, and $\sigma^{(n)}$
is the 
hard scattering cross section for partons in the final state,
calculated in perturbation theory\footnote{We have not indicated
technical complications such as different types of final-state
partons (quarks, gluons) and the convolution with parton densities
if there are hadrons in the initial state.}{}~. For the next-to-leading-order
calculation of the (2+1)-jet-production 
cross section to be described later on, the relevant 
values are~$n=2$ for the Born term and the virtual corrections, and~$n=3$ 
for the real corrections. Both virtual and real corrections exhibit 
infrared singularities. The soft singularities and the collinear singularities
in the final state cancel, owing to the Kinoshita--Lee--Nauenberg 
mechanism.
The collinear singularities from the initial state can be consistently 
absorbed into redefined parton densities. The resulting hard scattering 
cross section is finite. 

The factorization theorems of perturbative QCD 
(see Ref.~\cite{1} and references therein) 
show that parton cross sections~$\sigma$
have a very simple behaviour in collinear and soft phase space regions:
\begin{itemize}
\item In the collinear limit, for the partons with labels $j$ and $k$ being 
collinear, the limit for $\sigma$ is 
\beq
\label{cfac}
\sigma \sim \frac{\alpha_s}{2\pi}\,\frac{1}{s_{jk}}\,
\hat{P}_{kj\leftarrow i}(u)\,\sigma_{\mbox{\smc Born}},
\eeq
where~$u$ is the momentum fraction of the parent parton $i$ carried by 
parton~$j$, $s_{jk}=2p_jp_k$ and $\sigma_{\mbox{\smc Born}}$ is the 
Born cross section corresponding to the real correction $\sigma$
in the limit under consideration.
The $\hat{P}_{kj\leftarrow i}(u)$ are the unsubtracted Altarelli--Parisi
splitting functions (see, for example, Ref.~\cite{2}). 
There is a slight technical complication: factorization
in this form holds only for polarized cross sections. The sum over the 
parton polarizations introduces residual azimuthal correlations, even 
in the collinear limit. These have to be taken into account in the  
construction of the subtraction terms, see below. After integration over 
the azimuthal angle, factorization of the form of Eq.~(\ref{cfac}) holds
also in the unpolarized case.
The splitting functions $\hat{P}_{kj\leftarrow i}(u)$ are universal,
process-independent
functions. This has the important consequence that
physical parton densities and fragmentation functions can be defined in a 
process-independent way.
\item In the soft limit, the situation is slightly more 
complicated. For a gluon\footnote{Soft quarks do not lead to a
soft singularity. The matrix element is singular as $1/E$, but this
singularity is compensated by a factor~$E$ in the phase space volume.} 
{} with label $k$ becoming soft, the
factorization formula now reads
\beq
\label{sfac}
\sigma \sim \frac{\alpha_s}{2\pi}\,
\sum_{i, j \neq k}\,C_{ijk} \frac{p_i p_j}{(p_i p_k)(p_j p_k)}
\,\sigma_{\mbox{\smc Born}}.
\eeq
The sum runs over all pairs of partons $\{i, j\}$, and the $C_{ijk}$ are 
constant coefficients. The structure of this formula can be easily understood
in terms of an eikonal approximation, where the matrix element
$\cal M$ factorizes as 
\beq
{\cal M} \sim \frac{p_i \epsilon(\lambda)}{p_i p_k}\, 
{\cal M}_{\mbox{\smc Born}}.
\eeq
The sum over the gluon polarization $\lambda$, $\epsilon(\lambda)$ being the
gluon 
polarization vector, leads to the form given in Eq.~(\ref{sfac}).
\end{itemize}
In order not to spoil the cancellation of soft and collinear 
singularities from the real corrections against those of the virtual 
corrections, the observables $\cal O$ have to fulfill certain conditions.
This is the topic we now turn to.

An observable $\cal O$ is called {\it infrared-safe}, if the
functions ${\cal O}^{(n)}\left(p_1, \ldots, p_n\right)$ of the parton 
momenta respect soft and collinear limits:
\beqn
{\cal O}^{(n)}\left(p_1, \ldots, p_i, \ldots, p_n\right) 
&{{} \atop {{\displaystyle\longrightarrow}
             \atop 
            {\scriptstyle p_i \rightarrow 0}}}&
{\cal O}^{(n-1)}\left(p_1, \ldots, \hat{p}_i, \ldots, p_n\right),
\\
{\cal O}^{(n)}\left(p_1, \ldots, p_i, \ldots, p_j, \ldots, p_n\right) 
&{{} \atop {{\displaystyle\longrightarrow}
             \atop 
            {\scriptstyle p_i \parallel p_j}}}&
{\cal O}^{(n-1)}\left(p_1, \ldots, \hat{p}_i, \ldots, 
\hat{p}_j, \ldots, p_n, p_i + p_j\right).\nonumber
\eeqn
Momenta denoted by $\hat{p}$ 
are to be omitted.
The property of infrared safety for observables has the consequence
that the factorization from Eqs.~(\ref{cfac}) and~(\ref{sfac})
also works if $\sigma$ is replaced by $\sigma\,{\cal O}$.
This ensures the cancellation of the infrared singularities 
for the case of a convolution of the parton cross section with
an observable, as in Eq.~(\ref{thf}). 

\section{Universal Calculations in Next-to-Leading Order for Jet Quantities}
\label{CNLO}
The factorization theorems of perturbative QCD make it possible to 
perform calculations for a particular process in such a way that 
arbitrary infrared-safe observables can be evaluated numerically.
For this it is necessary to extract and cancel the infrared singularities 
in an observable-independent way. Technically, there are two different 
procedures to achieve this: the phase-space-slicing method \cite{3}
and the subtraction method \cite{4}. These techniques can be illustrated 
by means of a simple example \cite{5}\footnote{
Reference~\cite{5}
actually contains 
the first fully general algorithm to calculate subtraction terms.
}~. 
Assume that the integral
\beq
I=\int_0^A \dd x\, x^{-\epsilon} \,\frac{1}{x}\, f(x)
\eeq
is to be evaluated. The $x$-integration stands for the phase-space
integral, the factor $x^{-\epsilon}$ is the 
regulator\footnote{The calculations are performed in $d=4-2\epsilon$
space-time dimensions, so that UV and IR singularities are regulated and
made explicit in the form of poles in $\epsilon$. Infrared singularities
are regulated for $\epsilon<0$.}{}~, 
the term $1/x$ is the infrared 
singularity (a ``propagator term''), and the integrable 
function $f(x)$ contains regular parts of the cross section and of the 
observable. 
\begin{itemize}
\item In the case of the {\it phase-space-slicing method}, 
the range of integration is split into two parts $[0,a]$ and $[a,A]$
by means of an arbitrary small technical cut-off parameter~$a$.
For the lower interval, the function $f(x)$ can be approximated by $f(0)$
up to terms \porder{x}, which regulate the $1/x$ singularity, and, after
integration, give rise to terms \porder{a}. For the upper interval, 
the regulator $\epsilon$ can be set to zero. The final result for 
the integral is thus
\beq
I = -\frac{1}{\epsilon}f(0) + f(0) \ln a + \int_a^A\dd x\, \frac{1}{x}\, f(x)
+ \porder{a} +\porder{\epsilon}.
\eeq
The first term is the singular part, which would cancel against a similar
singular part of opposite sign from the virtual corrections. The remaining terms
are finite. The integral in the third term can be evaluated numerically.
The method is exact only in the limit $a\rightarrow 0$ (in practice, very small
values for $a$ are sufficient). The large logarithm in the second term 
is compensated by a similar logarithm from the numerical integration
in the third term.
\item The {\it subtraction method} requires the definition of a 
subtraction term, such that a splitting of the integral into a finite
and a singular part is possible. As an example we give a specific 
way to define the subtraction term:
\beq
I = \int_0^A\dd x\, x^{-\epsilon}\,\frac{1}{x}\, \left(f(x)-f(0)\right)
+\int_0^A\dd x\,x^{-\epsilon}\,\frac{1}{x}\,f(0).
\eeq
The subtraction (``local in phase space'') in the first integral renders
the integrand integrable for $\epsilon\rightarrow 0$, and the final 
result is thus
\beq
I = -\frac{1}{\epsilon}f(0) + f(0) \ln A + \int_0^A\dd x\, \frac{1}{x}\, 
\left(f(x)-f(0)\right)
+\porder{\epsilon}.
\eeq
The integral in the third term can again be performed numerically.
Because $f(0)$, the subtracted term in the integrand, is the residue
of $f(x)/x$ at $x=0$, this method is also called the {\it residue method}.
\end{itemize}
The advantage of the phase-space-slicing method is that it can be implemented
in a straightforward way: all that is required is the integral of 
the real corrections over small cones in phase space (corresponding to 
the approximation $x\approx 0$) and a numerical implementation 
in the form of a Monte-Carlo program of the
cross section itself. The disadvantage is that the cancellation 
of the term proportional to $\ln a$ is delicate; the statistical 
fluctuations require a large number of Monte-Carlo events.
Moreover, the convergence of the final result in the limit 
$a\rightarrow 0$ has to be checked, in principle, for every 
calculated expectation value. These technical problems are not present in 
the case of the subtraction method, because no small cut-off has to 
be used\footnote{Strictly speaking, a tiny cut-off of the
order of $10^{-8}$ to $10^{-10}$ has to be introduced in order to 
make the cancellation $\left(f(x)-f(0)\right)/x$ work, because 
there is only a finite number of significant digits available.}{}~. 
However, the
implementation of the method requires the construction of the
subtraction term, which is a non-trivial task. 
If this can be afforded, the subtraction method is 
the method of choice.

For the particular case of (1+1)- and (2+1)-jet production in deeply 
inelastic scattering, there are by now several calculations in the 
form of weighted Monte-Carlo programs available:
\begin{itemize}
\item {\tt PROJET} \cite{6}: The jet definition is restricted to 
the modified JADE scheme; the program is based on the calculation 
in Refs.~\cite{7,8,9}.
\item {\tt DISJET} \cite{10}: Again the jet definition is restricted to 
the modified JADE scheme; the program is based on the calculation 
in Refs.~\cite{11,12}.
\item {\tt MEPJET} \cite{13}: This is a program for the 
calculation of arbitrary
observables which uses the phase-space-slicing method. The calculation 
\cite{14} 
uses the Giele--Glover formalism \cite{15} 
for the analytical calculation of the
IR-singular integrals of the real corrections, and the crossing-function
technique \cite{16} 
to handle initial-state singularities. The latter requires
the calculation of ``crossing functions'' for each set of parton densities.
\item {\tt DISENT} \cite{17}: 
This program is based on the subtraction method. 
The subtraction term is defined by means of the dipole 
formalism\footnote{
The subtraction term is written as a sum over dipoles (an ``emitter'' formed
from two of the original partons and a ``spectator'' parton). Besides
the factorization theorems of perturbative QCD, the main ingredient is 
an exact factorization formula for the three-particle phase space, which allows
for a smooth mapping of an arbitrary 3-parton configuration onto the
various singular contributions.
}{}\,~\cite{18,19}. 
\item {\tt DISASTER++} \cite{20}: 
This is a {\tt C++} class library\footnote{
The acronym stands for ``Deeply Inelastic Scattering:
All Subtractions Through Evaluated Residues''.
Most of the program is written in {\tt C++}. A {\tt FORTRAN} interface
is in preparation; thus there will not be any problem to interface the
class library to existing {\tt FORTRAN} code.
}~{}.
The subtraction method
is employed, and the construction of the subtraction term resembles
the method of Ref.~\cite{5}, i.e.\ it is obtained by the evaluation
of the residues of the cross section in the soft and collinear limits.
Double counting of soft and collinear singularities
is avoided by means of a fully general 
partial fractions method.
\end{itemize}

Why a new calculation? There are two reasons: (a) The existing
programs have the restriction that the number of flavours is fixed 
($N_f=5$ in the case of {\tt MEPJET} 
and $N_f$ fixed, but arbitrary for {\tt DISENT}). 
For studies of the scale-dependence it is
necessary to 
have a variable number of flavours, 
in order to be consistent with the scale evolution 
of the strong coupling constant and the parton densities. 
{\tt DISASTER++} makes the $N_f$ dependence explicit in the ``user routine''
on an event-by-event basis,
and thus results for arbitrary renormalization and factorization scales
can be binned simultaneously.
(b) {\tt DISASTER++}
is already set up such that the extension to one-particle-inclusive 
processes will be possible without the necessity of re-coding 
the contributions which are already present for 
the jet-type observables. 

In the following we will illustrate how the subtraction term
for the real corrections is constructed. 
We will be concerned with infrared-safe observables that are zero
if more than one parton is soft, or if more than two partons are collinear, 
or combinations thereof.
These are the most relevant ones in practical applications.
In principle, the procedure is
simple. In energy and angle variables, soft singularities 
are of the form $1/E_i^2$, for parton $i$
being soft, and collinear singularities are given by terms $1/v_{ij}$, 
for partons $i$,~$j$ being collinear, where 
$v_{ij}=(1-\cos\vartheta_{ij})/2$, and $\vartheta_{ij}$ is the angle
between the two partons.
The singularities can be extracted by performing the limits
$E_i\rightarrow 0$ and $v_{ij}\rightarrow 0$ of the terms
$E_i^2 \sigma$, $v_{ij}  \sigma$ and $E_i^2 v_{ij} \sigma$, and the subtraction 
terms will consequently be given by
\beq
\frac{1}{E_i^2}\,\left[E_i^2\, \sigma\right]_{E_i=0}, \quad
\frac{1}{v_{ij}}\,\left[v_{ij}\, \sigma\right]_{v_{ij}=0}, \quad
\frac{1}{E_i^2 v_{ij}}\,\left[E_i^2\, v_{ij}\, 
\sigma\right]_{E_i=0,\,v_{ij}=0}.
\eeq
There is, however, the problem of overlapping singularities, or double 
counting. Consider a term of the form $1/(E_1^2\,v_{12}\,v_{13})$. Here
the problem is that if parton $1$ is soft, then there are two regions
which give rise to an additional collinear singularity: $v_{12}=0$
and $v_{13}=0$. In order to include every possible singular configuration
only once in the subtraction term, it is convenient to separate 
the singular terms. This can be done in two ways: ($\alpha$) division of the
phase space, or 
($\beta$) separation of the singularities in the matrix element.
In Ref.~\cite{5} a combination is used: $1/(v_{12}\,v_{13})$ is written 
in terms of partial fractions as
$1/[v_{12}(v_{12}+v_{13})] + 1/[v_{13}(v_{12}+v_{13})]$, 
and the energy of the $E_1$-integration
is restricted to be smaller than $E_2/2$. In principle, phase-space 
cuts of this kind do not pose a problem in Monte-Carlo programs. However, 
it is preferable that the integrand be a smooth function. We achieve
this by using the following general formula for partial fractions:
\beq
\label{pfid}
\frac{1}{x_1\,x_2\cdots x_n}
=\sum_{\sigma\in S_n}
\frac{1}{x_{\sigma_1}\,(x_{\sigma_1}+x_{\sigma_2})\cdots
         (x_{\sigma_1}+\ldots+x_{\sigma_n})}.
\eeq
The sum runs over all $n!$ permutations of $n$~objects\footnote{
This decomposition has a nice technical property. 
Holding $\sigma_1$ fixed, excluding $1/x_{\sigma_1}$
from the sum, and setting $x_{\sigma_1}$ to zero in the remainder, 
the sum over the
restricted set of permutations yields $1/(x_1\cdots\hat{x}_{\sigma_1}
\cdots x_n)$. This property is useful to recombine terms after the
soft and collinear limits have been performed.
}{}~. The most 
straightforward
way to apply this formula to the case at hand [(2+1)-jet production] would be 
to set the set of variables $\{x_1, \ldots, x_9\}$ to 
$\{E_1^2, E_2^2, E_3^2, 
v_{01}, v_{02}, v_{03}, v_{12}, v_{13}, v_{23}\}$ ($p_0$ is the momentum 
of the incident parton), where the energies are rescaled such that they 
are dimensionless. The product of the cross section and the 
observable $\sigma\,{\cal O}$ is then rewritten
as $[x_1\cdots x_n \,\sigma\,{\cal O}]/(x_1\cdots x_n)$, 
and the partial fractions
identity (\ref{pfid}) is applied to the denominator. The numerator is
regular: the two-parton singularities are regulated by the $x_i$, and the
remaining singularities are regulated 
by the observable (recall that we restrict the
discussion to this kind of observables).
The general partial fractions formula now gives us a ``hierarchy'' 
of singularities: the ``leading'' one $1/x_{\sigma_1}$, the ``subleading'' one
$1/(x_{\sigma_1}+x_{\sigma_2})$, and so forth. It turns out that for 
the observables under consideration, we have to consider only the leading
and subleading singularities. Let us write this singular part, for a specific
term in the sum, as $k_A$, 
and the remaining terms from the product $\sigma\,{\cal O}$ as 
$\tau_A$; the index $A$ parametrizes the set of permutations.
We then have $\sigma\,{\cal O} = \sum_A k_A \tau_A$. Each of the terms 
is singular only if the ``leading variable'' is zero. The subtraction
term can therefore
be constructed with respect to the leading and subleading 
variable in $k_A$. More precisely this is done in the following way.
For each index $A$ there is a particular parton label $i_A$ related to it
(for example, for
$1/[E_1^2\,(E_1^2+v_{12})]$ or $1/[v_{12}\,(E_1^2+v_{12})]$ this would 
be~$i_A=1$). The phase space $\dps^{(n)}$ is factorized according to
$\dps_{i_A}\,\dps^{(n-1)}$, i.e.\ the one-parton
phase space corresponding to parton $i_A$ is pulled out of the
full phase space. Finally the phase space integration from Eqn.~(\ref{thf})
is rewritten
as
\beqn
\int\dps^{(n)}\,\sigma\,{\cal O}
&=& \sum_A \int \dps_{i_A} \,k_A \left(
\int \dps^{(n-1)} \tau_A - 
 \left[
   \int \dps^{(n-1)} \tau_A
 \right]_{\mbox{\smc soft/coll.~limit}}
\right)\nonumber\\
&+& \sum_A \int \dps_{i_A} \,k_A \left[
   \int \dps^{(n-1)} \tau_A
    \right]_{\mbox{\smc soft/coll.~limit}}.
\eeqn
The first integral is finite and can be calculated numerically. The second 
integral contains all infrared singularities. The term in the square bracket 
is simple because of the factorization theorems of QCD, and the one-particle
integral over the kernel $k_A$ and the factorization contribution from the
term in the square brackets can be performed easily\footnote{There are a few
complications, though. The first one is that the required sum of $9!$ 
terms, which has to be done numerically, contains too many terms to be 
efficient (or even possible). The solution is to perform the partial fractions
decomposition separately for the energy and angle terms. 
The second complication is the presence of correlations in azimuthal angle
for collinear singularities, which have been mentioned above. They can
be dealt with by introducing a fictitious azimuthal variable for
the collinear configurations.}{}~.

\section{Numerical Results and 
Applications: {\boldmath$\alpha_s$} and the Gluon Density}
\label{ANLO}
We now turn to some applications of (2+1)-jet observables.
Loosely speaking, observed jets of hadrons can be identified
with partons at hard scales. The subsequent fragmentation process is assumed
to be sufficiently soft in order to keep non-perturbative effects
small. Jets cannot be defined in a canonical way, rather they are
objects by definition. It is convenient to define jets in terms of 
iterative cluster algorithms\footnote{
``Non-cluster'' algorithms of the cone-type suffer from 
the problem that the outcome may depend on the choice of
a seed in the jet-finding process. This effect is particularly large
at small transverse momentum. Cluster algorithms are well-defined
without any ambiguity in the procedure.}~. There are three ingredients
in the definition of a cluster algorithm: (i) a distant measure which 
determines the relative distance of two clusters in momentum space
(for example their invariant mass $s_{ij}=(p_i+p_j)^2$, (ii) a mass scale $M^2$
that determines whether two clusters are to be combined into a single one
(for example if $s_{ij}<M^2$), and (iii) a recombination procedure
that prescribes how two clusters~$p_i$ and~$p_j$ are to be merged into a 
single cluster~$p_*$
(for instance $p_* = p_i + p_j$). 
There are various choices
for (i), (ii) and (iii). The example given is the JADE algorithm
\cite{21} in the so-called E-scheme. This particular algorithm is known for 
large hadronization corrections. This problem can be
reduced 
by explicitly forcing clusters to behave as if they were massless 
(JADE and P-schemes). The large corrections partly come
from the fact that for JADE-type algorithms, soft partons are combined
first, even if they differ considerably in their direction. Intuitively, 
such partons should possibly be combined with other partons nearby, even
if they are hard. This can be achieved by means of a distance measure 
based on relative transverse momentum, leading to the 
$k_T$~algorithm, defined in the Breit frame of reference. 
For deeply inelastic scattering, it 
has been introduced in Ref.~\cite{22}\footnote{
These algorithms have the property of factorization, which means that 
the $(n+1)$-jet cross section factorizes in exactly the same way into 
a product of a coefficient function and a parton density as a
structure function does, without any residual explicit $x_B$-dependence.
This kind of factorization is technically convenient for the calculation
of resummed cross sections. We wish to stress that any acceptance cut
on the final-state jets destroys this factorization property.
What is important to have in a well-defined perturbative prediction is that
the {\it singular} 
parts factorize in a universal way, which permits the definition
of process-independent parton densities. This is, of course, the case 
for all infrared-safe observables, including the non-factorizing kind
of jet algorithms.}~.
 
In the case of hadrons in the initial state, the corresponding remnants
need a special treatment, because partons in the very forward direction
lead to initial-state singularities. This problem is taken care of by 
either including the remnant jet in the clustering procedure (as in the 
case of the {\it modified} JADE algorithm) 
or by discarding partons which are
too close to the remnant (as for the $k_T$~algorithm). In any case
a well-defined prediction for the (2+1)-jet cross section is obtained.

\begin{figure}[htbp] \unitlength 1mm
\begin{center}
\dgpicture{160}{72}
\put(15,10){\epsfig{file=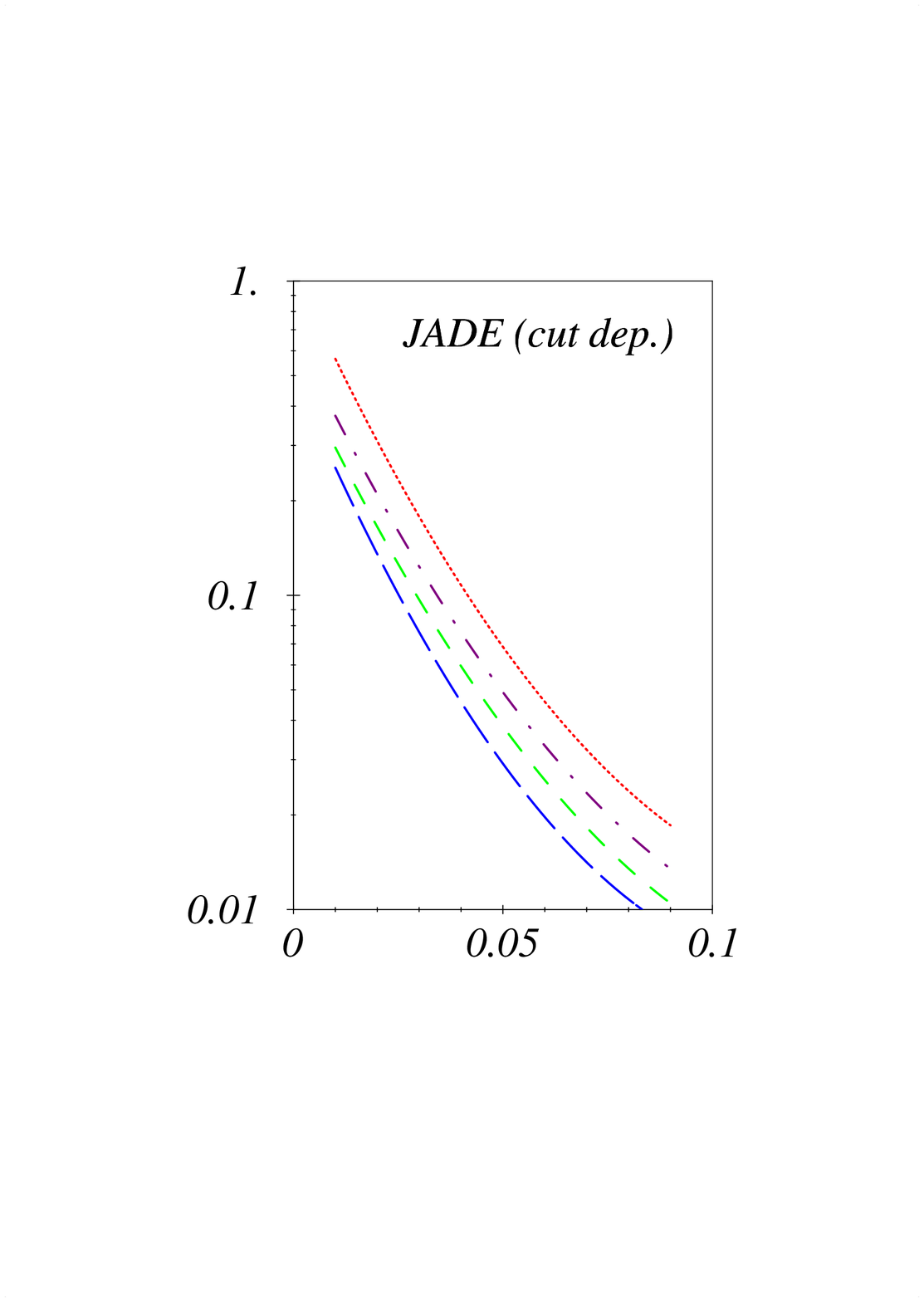,width=4.5cm}}
\put(90,10){\epsfig{file=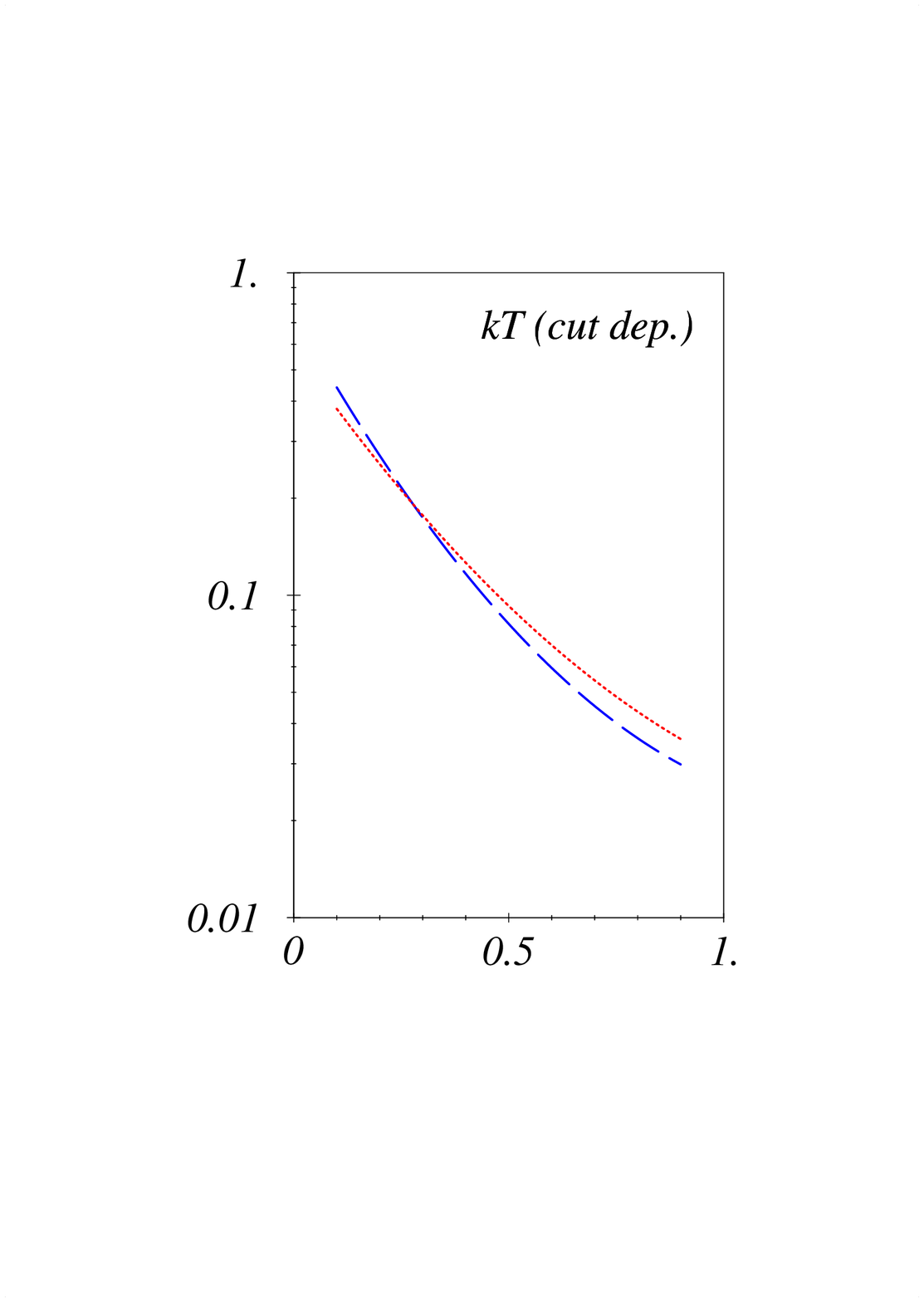,width=4.4cm}}
\put(15, 0){(a)}
\put(90, 0){(b)}
\put( 57, 4){$\scriptstyle c$}
\put(132, 4){$\scriptstyle c$}
\put(18, 70){$\scriptstyle R_{(2+1)}$}
\put(92, 70){$\scriptstyle R_{(2+1)}$}
\end{picture}
\end{center}
\shiftcaption
\caption[]
{{\it Cut-dependence of $R_{(2+1)}$ for the modified JADE cluster algorithm
(a) and for the $k_T$~algorithm (b). The jet resolution mass is
$M^2=c W^2$ for the JADE algorithm and $M^2=c Q^2$ for the $k_T$~algorithm.
Leading order [\longdashline], next-to-leading order:
E-scheme \mbox{[\dotline]}, 
P-scheme \mbox{[\dashline]}, 
JADE-scheme \mbox{[\dashdotline]}.
The recombination scheme dependence of the $k_T$~algorithm is small, 
therefore only the E-scheme is shown in this case.
}}
\label{figref01}
\end{figure}

In order to reduce systematic errors, it is convenient to consider
the (2+1)-jet rate $R_{(2+1)}=\sigma_{2+1}/\sigma_{\mbox{\smc tot}}$.
Figure~\ref{figref01} shows the cut-dependence of $R_{(2+1)}$ for
two different jet algorithms\footnote{ 
{}\,\,The centre-of-mass energy is 300~GeV, 
the photon virtuality is restricted by 
$(10\,\mbox{GeV})^2\,<Q^2<\,(20\,\mbox{GeV})^2$, the total hadronic
energy $W$ is assumed to be larger than 70~GeV 
and the lepton variable~$y$ is restricted to values smaller than~0.7.
The parton densities are MRS~A$^\prime$ \cite{23}, 
and the renormalization and
factorization scales are set to $Q^2$.}{}~. The recombination scheme dependence
of the modified JADE algorithm is large, as shown in Fig.~\ref{figref01}a. 
For the E-scheme the QCD corrections may be as large as 100\,\%; 
they are considerably smaller for the ``massless'' recombination 
schemes\footnote{
\,The large recombination scheme dependence of the JADE algorithm 
sometimes leads to the conclusion that the algorithm is
ill-defined. This is a misconception. Different
recombination schemes define different observables, and there is no
reason why different observables should not lead to different
predictions. However, 
the strong dependence of the jet rate on the recombination scheme
points to the fact that jet masses are actually important.
}{}~.
The perturbative stability of the $k_T$~algorithm is better, here the
QCD corrections are of the order of a few per cent (Fig.~\ref{figref01}b).

\begin{figure}[htbp] \unitlength 1mm
\begin{center}
\dgpicture{160}{76}
\put( 0,10){\epsfig{file=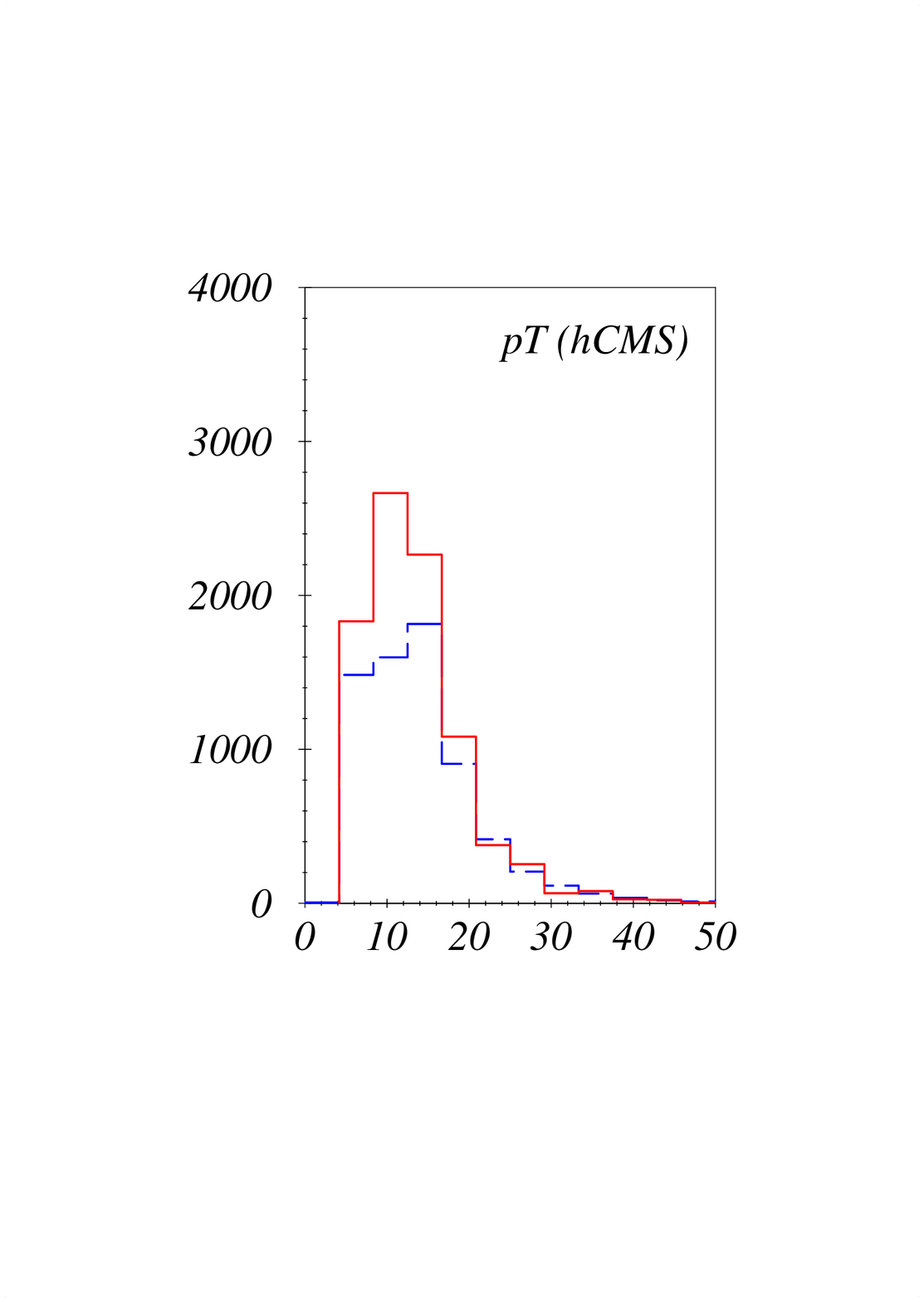,width=4.5cm}}
\put(50,10){\epsfig{file=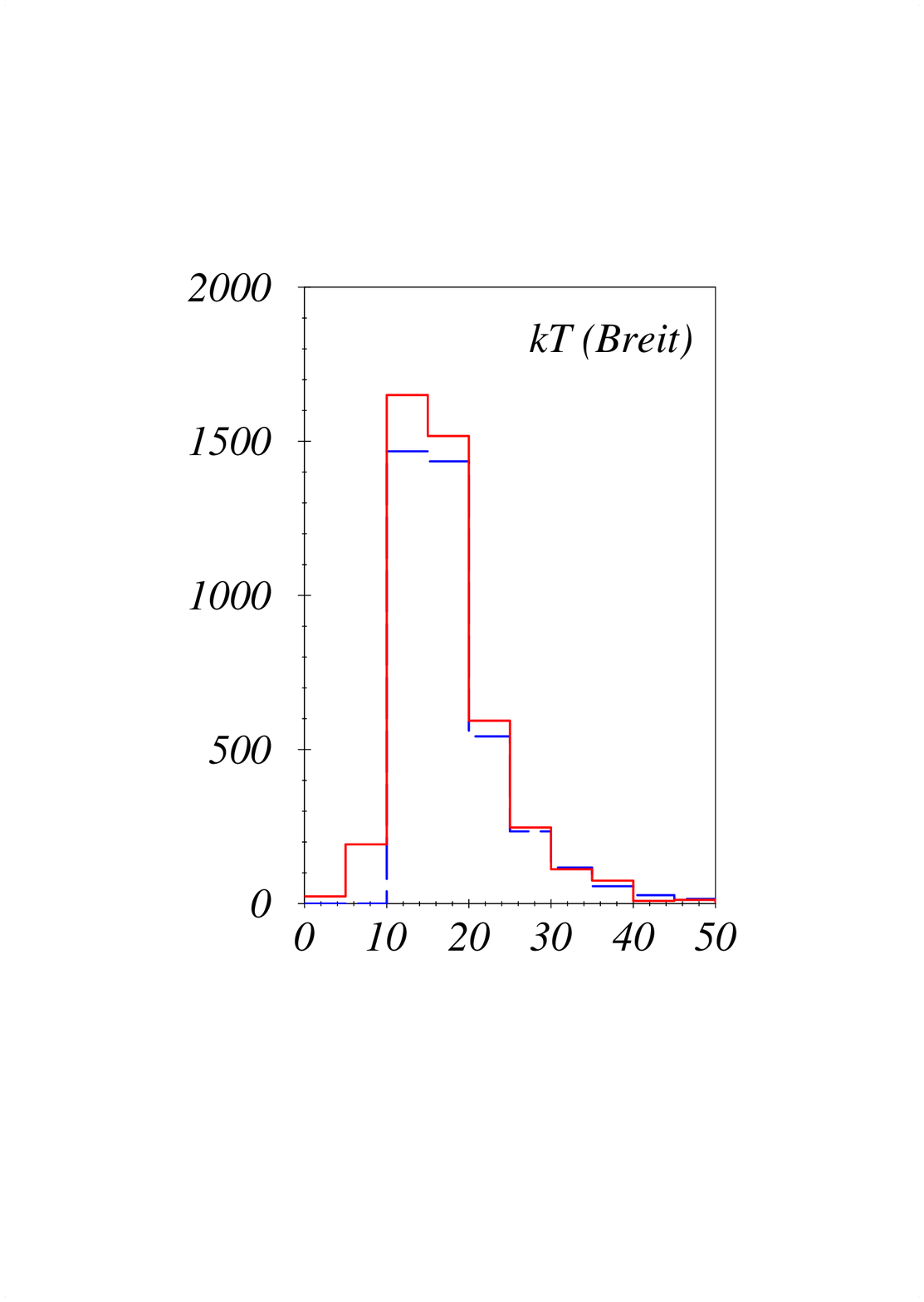,width=4.5cm}}
\put( 0, 0){(a)}
\put(50, 0){(b)}
\put( 42, 4){$\textstyle p_T$}
\put( 92, 4){$\textstyle p_T$}
\put(-1, 72){
   $\textstyle\frac{{\dds} 
                       \sigma [
                       \mbox{{\scriptsize pb}} 
                       ]}
                      {\dds x_B \,\dds y\, \dds p_T [\mbox{\scriptsize GeV}]}\,
   $}
\put(49, 72){
   $\textstyle\frac{\dds \sigma [\mbox{\scriptsize pb}]}
                      {\dds x_B \,\dds y\, \dds p_T [\mbox{\scriptsize GeV}]}\,
   $}
\put(105,10){\epsfig{file=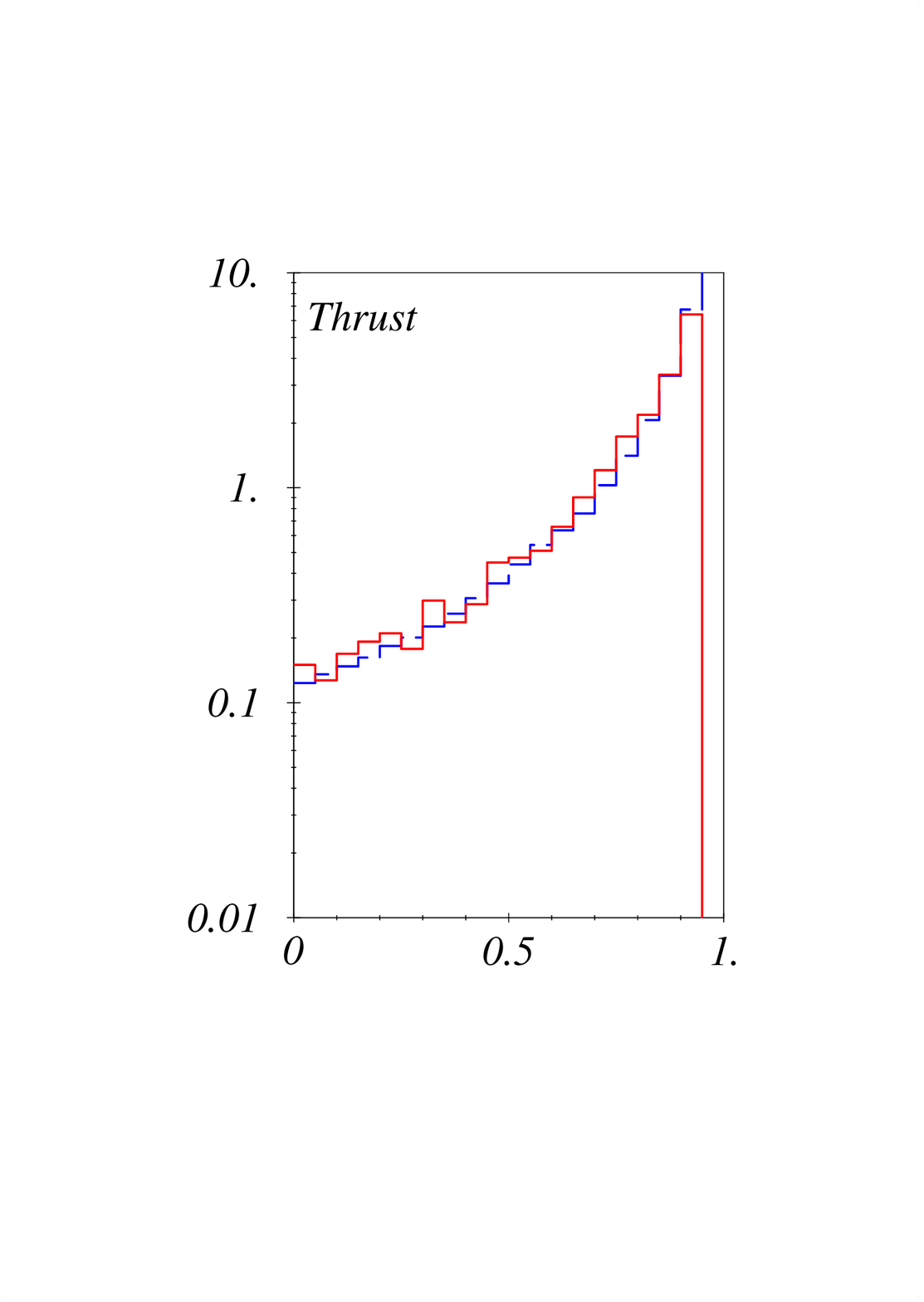,width=4.3cm}}
\put(105, 0){(c)}
\put(145, 4){$\textstyle T_z$}
\put(106, 72){$\textstyle \frac{1}{\sigma_{\mbox{\tiny tot}}}\,
               \frac{\dds \sigma}{\dds T_z}$}
\end{picture}
\end{center}
\shiftcaption
\caption[]
{{\it Transverse momentum distribution
in the hadronic CMS for the JADE algorithm (a), and in 
the Breit system for the $k_T$ algorithm (b).
Thrust distribution in the Breit frame (c).
Shown are the leading order \mbox{[\longdashline]} and the 
next-to-leading order \mbox{[\fullline]}.
}}
\label{figref02}
\end{figure}

Deeply inelastic scattering is a multi-scale problem. Besides the photon 
virtuality $Q^2$ there are other hard scales related to the
scattering process, for example the transverse momentum~$p_T$
of partons or jets.
These hard scales can be very different from each other.
The transverse momentum distribution of the jets for fixed 
lepton variables of $x_B=0.01$ and $Q^2=(20\,\mbox{GeV})^2$
is shown in Fig.~\ref{figref02}
for the JADE P-scheme with a recombination scale of $0.02\,W^2$
in the hadronic CMS~(a) and for the $k_T$ algorithm 
with a recombination scale of $0.5\,Q^2$ in the Breit frame~(b).
The transverse momentum distribution is quite broad, and therefore~$p_T^2$ 
is not 
necessarily of the order of~$Q^2$. This introduces 
a theoretical uncertainty, because it is not clear whether $Q^2$ or
$p_T^2$ is the right scale to be used in the renormalized coupling
$\alpha_s(\mu_r^2)$ and in the parton densities $f_i(\xi,\mu_f^2)$.
As long as no resummed calculation is available, this uncertainty
has to be considered as a contribution to the systematic error in 
case that physical quantities are extracted by using fixed-order
next-to-leading-order calculations\footnote{Not shown here because of
lack of space is a comparison of the scale dependence in leading 
and next-to-leading order. For the jet cross sections shown in 
Fig.~\ref{figref01} the scale dependence is reduced by about a factor
of~2, if the renormalization and factorization scales are varied
between $0.5\,Q$ and $2\,Q$.}{}~. 

As an example for an event shape variable we show a Thrust distribution
in Fig.~\ref{figref02}c. The phase space region is the same as
the one of the jet rates shown before.
The Current Thrust variable is defined by
$T_z=2\sum_i p_{i\parallel}/Q$
\cite{24}, where the sum is over 
all partons in the current hemisphere (the hemisphere 
containing the virtual photon). The quantity $p_{i\parallel}$ is
the longitudinal 
component of the parton momentum $p_i$
along the virtual photon direction in 
the Breit frame of reference. The Thrust variable goes to one for a 
(1+1)-jet-like configuration. Values of $T_z < 1$
are therefore obtained from processes
of \porder{\alpha_s}. The QCD corrections are moderate.
It turns out that the NLO calculations do not describe the experimental 
data well (see, for example, Ref.~\cite{25}). 
The inclusion of a power suppressed term
of the form $A/Q$ with a large coefficient $A$ brings data and the theoretical
prediction in good agreement. A quantitative derivation for the coefficient
$A$ in terms of an effective coupling constant at low scales is given in 
Ref.~\cite{26}.

Applications of NLO calculations in deeply inelastic scattering include 
the measurement of the strong coupling constant
\cite{27,28} and a direct fit of the gluon density \cite{29}.
The $\alpha_s$~measurements show a clear evidence for the running of 
the coupling constant as a function of the scale $Q^2$.
To complement the measurement via jet rates, measurements of 
$\alpha_s$ based on event shape variables are currently under way
\cite{25}\footnote{
For the Thrust variable, the requirement of a power-suppressed term $A/Q$ 
makes an $\alpha_s$~measurement conceptually difficult, because the
coefficient~$A$ is parametrized by a universal effective coupling
constant~$\alpha_{s,\mbox{\smc eff.}}$, 
whereas the perturbative contribution has the renormalized coupling constant
as a coefficient. There are matching procedures to disentangle the two 
terms, but the size (about the same order of magnitude as the perturbative
contribution) of the non-perturbative contribution is disturbing.
}{}~. 
The gluon density has been determined via jet rates by means of a
Mellin transform method \cite{30,31} which makes the repeated 
evaluation of the NLO cross section for the purpose of the 
fitting procedure feasible\footnote{
This method is actually quite general and permits the calculation
of arbitrary observables, the only limitation being that 
the factorization scale has to be fixed.}~. 
The quark distributions are well known, 
and therefore the quark-initiated contribution can be subtracted.
The obtained direct fit is in good agreement with gluon density 
parametrizations from global fits\footnote{
For a specific value of the momentum fraction, the error bands
show a cross-over. This phenomenon is well known from global fits
\cite{32} and is a consequence of the limited number of 
parameters in the {\it ansatz} for the gluon density.
}{}~.

\section{One-Particle-Inclusive Processes}
\label{OPIP}
One-particle-inclusive processes are another very promising field.
The corresponding cross sections can be considered to be related 
to a special type of observable ${\cal O}$, the fragmentation functions
$D_i\left(\xi,\mu_D^2\right)$. 
In the terminology used above, they are not infrared-safe, so that
collinear singularities remain, which do not cancel. However, the 
singularities are of a universal form, and can be absorbed into the
fragmentation functions. The redefined fragmentation functions are 
finite and universal, i.e.\ process-independent.

\begin{figure}[htb] \unitlength 1mm
\begin{center}
\dgpicture{160}{94}
\put(30,-3){\epsfig{file=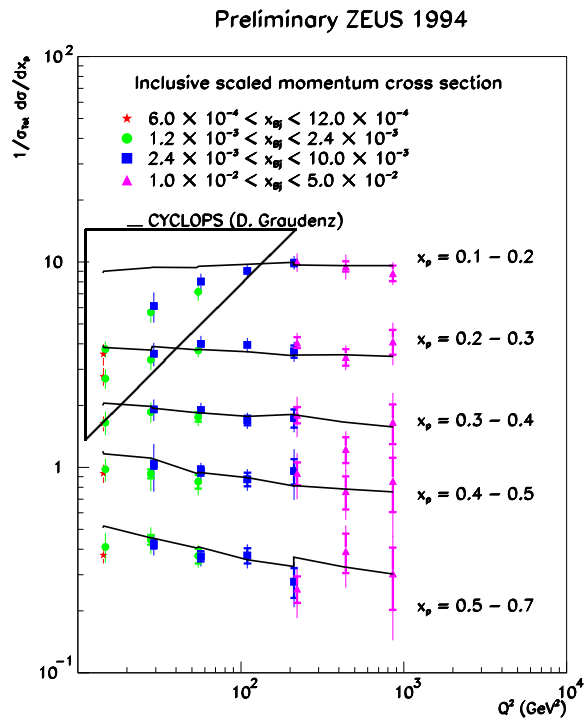,width=10cm}}
\thicklines
\put(46,37){\line(1,1){28}}
\put(46,65){\line(1,0){28}}
\put(46,65){\line(0,-1){28}}
\end{picture}
\end{center}
\shiftcaption
\caption[]
{{\it Theoretical prediction vs.\ experimental data for the  
$Q^2$-dependence of
$\sigma^h/\sigma_{\mbox{\smc tot}}$ for various bins of $x_p$.
The kinks in the theoretical curves come from the overlay of several bins
in~$x_B$ for the same bin in~$Q^2$. The region within the triangle
cannot be described by means of the fragmentation function picture; see text.
}}
\label{figref03}
\end{figure}

Figure~\ref{figref03} shows a comparison of recent preliminary experimental 
data for charged hadron production
from the ZEUS collaboration with a next-to-leading order calculation.
The employed variable is $x_p=2P_h/Q$, where $P_h$ is the momentum of 
the observed charged hadron~$h$ in the current hemisphere of the Breit frame.
The cross section~$\sigma^h$ 
is a convolution of parton densities, fragmentation
functions and perturbatively calculated
coefficient functions. The parton densities are from the
MRS~A$^{\prime}$ set and the fragmentation functions are from 
Refs.~\cite{33,34}. The coefficient functions have been calculated 
in Ref.~\cite{35}. The numerical results have been obtained by using 
the implementation in the program {\tt CYCLOPS} of a recent recalculation
\cite{36}. The comparison of the theoretical prediction with experimental
data shows a very good agreement, except for small values of $Q^2$ and  
small values of $x_p$. 

What is the reason for this discrepancy? It is instructive to 
consider the situation in terms of the rapidity variable~$y_h$
of the observed hadron. We consider 
the leading-order process, with a single parton fragmenting into the
hadronic final state. Under the assumption that the 
hadron has a typical
transverse mass~$m_T$ of \porder{500\,\mbox{MeV}}, the relation between
$x_p$ and~$y_h$ can be calculated. In the Breit frame, a positive value
of~$y_h$ stands for production in the current direction, and a negative
value for production in the target direction. In order to make the 
fragmentation function picture a valid description, the hadron 
should be produced closely in rapidity to the parent parton, which means 
that~$y_h$ should be larger than about $y_{\mbox{\smc min}}=1$ units 
of rapidity.
This can be translated into a lower bound on $x_p$:
\beq
x_{p,\mbox{\smc min}} = \frac{2m_T}{Q}\,
  \frac{1}{\sqrt{1-\left(\tanh y_{\mbox{\smc min}}\right)^2}}. 
\eeq
This excludes the region roughly indicated by the triangle in the figure.
The theoretical prediction thus fails in this region because the
fragmentation function picture is not applicable: the hadron is
not produced sufficiently close to the parent parton.
Another reason is that mass effects (which are not included
in the fragmentation function formalism) 
become important at small $Q^2$ and
small $x_p$ for $x_p\approx 2m_\pi/Q$. This excludes a similar region
in the plot. Outside of this critical region, in particular at large
values of $Q^2$, the good agreement shows the universality of 
fragmentation functions.

A future application might be the determination of $\alpha_s$ via
scaling violations of the fragmentation functions. This has been studied 
in some detail in Ref.~\cite{37}. It turns out that the main 
uncertainty is the dependence on a parametrization of the parton densities.
This dependence is reduced at large values of $Q^2$, but then the
available luminosity is the limiting factor. Another possibility might
be the measurement of $\alpha_s$ from the hard scattering matrix element
by means of $p_T$~spectra of charged particles. For this the NLO
calculation is not yet available; however, as indicated above, 
the implementation of the matrix elements in {\tt DISASTER++} is already 
such that it may become available in the near future. The advantage 
of such a measurement would be that the fragmentation process is no longer
modelled by means of event generators as in the case of jet cross sections
or event shape variables, but described by a universal parametrization
of fragmentation functions with a QCD-predicted scale evolution.

\xsection{Summary and Outlook}
We have reviewed the status and the applications of next-to-leading-order
calculations in deeply inelastic lepton--nucleon scattering. In the 
last few years, 
programs have become available that permit the calculation of arbitrary
infrared-safe observables of the (2+1)-jet type. In next-to-leading order, 
these programs are still restricted to the case of photon exchange. A natural
extension would be the case of $Z$~exchange, and the inclusion of 
charged-current processes. Moreover, the next-to-leading order is not
yet available for polarized initial states, and for the $p_T$~spectra
of identified hadrons. 

Concerning the phenomenology of hadronic final states, the main problem is
to understand the phase space region of small $Q^2$ (where, for example, 
the Thrust distribution is not well described in next-to-leading order), 
and the region of forward jets. With high-statistics data, the extraction of
physical quantities can be improved by hard cuts to remove these
dangerous regions, but a better theoretical understanding is certainly 
desirable. The physics of deeply inelastic hadronic final states will 
thus continue to be a very interesting topic in the future.

\xsection{Acknowledgements}
I would like to thank the organizers of the workshop for
having prepared a very interesting and stimulating meeting.
Moreover I gratefully acknowledge 
discussions with J.~Bromley, N.~Brook, J.~Collins, 
T.~Doyle, Z.~Kunszt and T.~Sj\"{o}strand.
The ZEUS Collaboration provided me with the plot from Fig.~\ref{figref03}.

\xsection{References}

\newcommand{\bibitema}[1]{\bibitem{#1}}

\end{document}